\documentclass{article}
\usepackage[a4paper, left=1.5cm, right=1.5cm, top=2.5cm, bottom=2.5cm]{geometry}
\usepackage{graphicx} 
\usepackage{amsmath}
\usepackage{amssymb}
\usepackage{xcolor}
\usepackage{authblk} 
\usepackage{cite}

\title{A Comprehensive Network for the Discovery and Characterization of Interstellar Objects}

\author[1]{Oem Trivedi\thanks{oem.trivedi@vanderbilt.edu}}
\author[2]{Abraham Loeb\thanks{aloeb@cfa.harvard.edu}}

\affil[1]{Department of Physics and Astronomy, Vanderbilt University, Nashville, TN 37235, USA}
\affil[2]{Astronomy Department, Harvard University, 60 Garden St., Cambridge, MA 02138, USA}

\date{\today}

\begin{document}

\maketitle

\begin{abstract}
Interstellar object (ISO) astronomy has rapidly emerged over the past decade as a new frontier in planetary astrophysics, yet current observations remain limited by short visibility windows, inference degeneracies and fragmented follow-up capabilities. We argue that these constraints are structural rather than incidental and motivate a coordinated, end-to-end observational strategy for future ISO studies. We propose the Comprehensive ISO Network (CISON) which combines dual hemisphere wide-field discovery with rapid high resolution characterization and selective escalation to interceptor missions. By coupling this architecture to the differential formulation of the Loeb Scale, ISO classification and risk assessment become predictive rather than reactive. This framework transforms ISO astronomy into a mature, scalable discipline capable of maximizing scientific return and informing planetary defense in the coming decades.
\end{abstract}

\section{Introduction}

The last decade has ushered in the discovery of Interstellar Objects (ISO), marking the emergence of a genuinely new observational window onto planetary astrophysics beyond the Solar System. The discoveries of 1I/‘Oumuamua \cite{oumumeech2017brief}, 2I/Borisov \cite{borguzik2020initial}, and most recently 3I/ATLAS \cite{seligman2025discovery} have demonstrated unambiguously that the Solar System is not an isolated environment but is instead permeated by a flux of interstellar material originating from other planetary systems. These detections have provided the first direct empirical probes of extrasolar planetesimals offering insights into the formation, composition and dynamical processing of small bodies in environments far removed from our own \cite{oum1loeb2022possibility,oum2forbes2019turning,oum3siraj2022mass,oum4siraj20192019,oum5bialy2018could,3i1loeb20253i,3i2a2hibberd2025interstellar,3i3tde2025assessing,3i4a3loeb2025intercepting,3i5hopkins2025different,ref2scarmato2026rotation,ref3barbieri2026rare}. In doing so, ISO astronomy has begun to connect planetary science, stellar dynamics and Galactic astrophysics in a way that was previously accessible only through indirect inference \cite{iso5siraj2022new,iso1seligman2022interstellar}.
\\
\\
At the same time, the rapid progress of the field has highlighted how young and structurally incomplete ISO astronomy remains \cite{iso3hajdukova2020challenge}. Despite notable successes, current discoveries are rare, observationally constrained and often characterized by substantial degeneracies in physical interpretation, and detection is limited by short visibility windows and survey cadence. Beyond this, follow-up observations are frequently reactive, fragmented and constrained by atmospheric or scheduling limitations \cite{iso2herbst2001chemistry,iso4jewitt2023interstellar,iso6guelin2022organic}. As a result, many of the most fundamental questions regarding ISO size, shape, composition, internal structure and dynamical history, remain weakly constrained. These challenges do not undermine the promise of the field, but rather emphasize that the present era represents an early, exploratory phase in which observational capability has outpaced the development of a coherent end-to-end strategy \cite{iso7mann2010interstellar,iso8savage1979observed,iso9aannestad1973interstellar}.
\\
\\
This situation naturally motivates the need to think beyond individual facilities and toward a future observational architecture for ISO astronomy that can scale with rising discovery rates and increasing scientific and societal relevance. In this work we propose such a framework, designed to integrate discovery, rapid characterization and quantitative decision making into a unified system and our work is organized as follows. In Section 2 we examine the principal limitations of current ISO observational capabilities and identify the structural issues that constrain progress. In Section 3, we introduce a novel coordinated architecture for future ISO astronomy and discuss its feasibility and advantages. In Section 4 we illustrate the power of this approach through a hypothetical case study of a newly discovered interstellar object and in Section 5 we summarize our conclusions and outline the broader implications for the future of the field.
\section{Current Limits of ISO Astronomy}

Despite the rapid growth of time domain surveys, the current practice of ISO astronomy remains fundamentally constrained by a small number of structural limitations that collectively restrict both the rate of discovery and more importantly, the depth of physical inference that can be drawn from any individual detection. These limitations do not arise from a lack of observational effort in any case but from the intrinsic mismatch between the transient, fast moving nature of ISOs and the capabilities of existing discovery and follow-up infrastructure. In what follows, we focus on four dominant issues that presently define the boundary of what ISO astronomy can achieve.
\\
\\
A first and primary limitation is that ISO discovery is inherently an $\acute{e}$cadence limited problem, while the visibility window of an ISO is intrinsically short. The apparent magnitude of an object with absolute magnitude $H$ observed at heliocentric distance $r$ and geocentric distance $\Delta$ can be written as
\begin{equation} \label{mag}
m \simeq H + 5\log_{10}(r\Delta) + \Phi(\alpha),
\end{equation}
where $\Phi(\alpha)$ is a phase angle dependent correction and because ISOs are typically detected at large phase angles and on hyperbolic trajectories with characteristic velocities of tens of km s$^{-1}$, the time during which $m$ remains below a survey limiting magnitude $m_{\rm lim}$ is often of order weeks to months at most. The characteristic visibility time can be approximated as
\begin{equation}
t_{\rm vis} \sim \frac{2R_{\rm vis}}{v},
\end{equation}
with $R_{\rm vis}$ being the heliocentric distance at which the object remains detectable and $v$ is its heliocentric speed. Any incompleteness in sky coverage, cadence gaps or delays in alert dissemination directly translate into missed discoveries. This places an irreducible premium on wide-field, high-cadence surveys and simultaneously limits the number of ISOs that can be discovered with sufficient warning for meaningful follow-up.
\\
\\
A second major limitation arises after detection, with the severe degeneracy in photometric and astrometric inference caused by short observational arcs and unfavorable viewing geometry. The reflected flux from an ISO scales as
\begin{equation}
F \propto \frac{p\,A_{\rm eff}}{r^2\Delta^2}\,\Psi(\alpha),
\end{equation}
with $p$ being the geometric albedo, $A_{\rm eff}$ being the effective cross-sectional area and $\Psi(\alpha)$ encoding the scattering phase function. Photometric data constrain only the product $pA_{\rm eff}$ convolved with poorly known scattering behavior, leaving size and albedo strongly degenerate. While light curve periodicity may suggest elongation, for a rotating body with 
\begin{equation}
\Delta m \simeq 2.5\log_{10}\!\left(\frac{a}{b}\right),
\end{equation}
this relation depends sensitively on viewing angle and surface heterogeneity, which are both poorly constrained by sparse cadence data. At the same time, orbital inference based on short astrometric arcs leads to inflated uncertainties in the inbound asymptote and hyperbolic excess velocity, limiting the ability to associate ISOs with Galactic source populations \cite{iso1seligman2022interstellar,iso7mann2010interstellar}. These degeneracies mean that even basic physical quantities such as size, shape and origin remain weakly constrained for most detected objects.
\\
\\
A third limiting factor is the ambiguity in interpreting non-gravitational accelerations, which has emerged as a central theme in ISO studies. If the equation of motion is written as a sum of gravitational and non-gravitational components
\begin{equation}
\ddot{\mathbf{r}} = \ddot{\mathbf{r}}_{\rm grav} + \mathbf{a}_{\rm ng},
\end{equation}
then the physical origin of $\mathbf{a}_{\rm ng}$ is generally unknown and in the case of outgassing, one expects
\begin{equation} \label{ng}
a_{\rm ng} \sim \frac{f\dot{m}u}{M},
\end{equation}
where $\dot{m}$ is the mass loss rate with an excess fraction $f$ in a preferred direction, $u$ the exhaust speed and $M$ the body mass. In a radiation pressure interpretation, the scaling becomes
\begin{equation}
a_{\rm rad} \sim \frac{P_{\rm rad} A}{M},
\end{equation}
with $P_{\rm rad} = \frac{L_0}{4 \pi r^2 c}$ being the solar radiation pressure and $A/M$ is the projected area-to-mass ratio. It is important to emphasize that both interpretations hinge on combinations of physical parameters that are not independently measurable with current data products. The absence of a detected coma does not uniquely rule out outgassing, while radiation-pressure modelling require extreme values of $A/M$. As a result, non-gravitational accelerations introduce a deep and persistent degeneracy that cannot be resolved with photometry and astrometry alone, first exemplified in the case of 1I/'Oumuamua \cite{iso4jewitt2023interstellar}.
\\
\\
The fourth and arguably most fundamental limitation is the lack of spatial resolution and rapid response follow-up. For a diffraction-limited system of effective baseline $D$, operating at wavelength $\lambda$, the angular resolution is
\begin{equation}
\theta \sim \frac{\lambda}{D},
\end{equation}
corresponding to a linear resolution $\ell \sim \Delta\theta$ at distance $\Delta$. Resolving kilometer scale structure at $\Delta\sim 1$ AU requires
\begin{equation} \label{kmscale}
D \sim \frac{\lambda \Delta}{\ell} \sim 10^2\ {\rm m},
\end{equation}
for optical wavelengths, far beyond the capabilities of current ground-based facilities in an operationally responsive mode. Consequently, morphology and surface structure remain inaccessible as well and spectroscopy is often photon-starved due to the rapid fading of the object as $\Delta$ increases. Because the photon yield scales as the time integral of the flux, delays in coordinated follow-up lead to a disproportionate loss of information, which ends up making ISO science intrinsically short-lived under current operational paradigms.
\\
\\
Taken together, these four issues show the present limitations of ISO astronomy. The big picture that is being painted here is that discovery is limited by cadence and visibility windows, physical inference is dominated by photometric and dynamical degeneracies, non-gravitational effects remain fundamentally ambiguous and the absence of rapid, high-resolution characterization prevents the collapse of these degeneracies. These limitations are not independent but mutually reinforcing. They motivate the need for an observational architecture that explicitly separates and optimizes discovery and characterization, while preserving information content through rapid response and access to fundamentally new measurement modes.

\section{The Comprehensive InterStellar Objects Network (CISON)}

\begin{figure}[!h]
    \centering
    \includegraphics[width=1\linewidth]{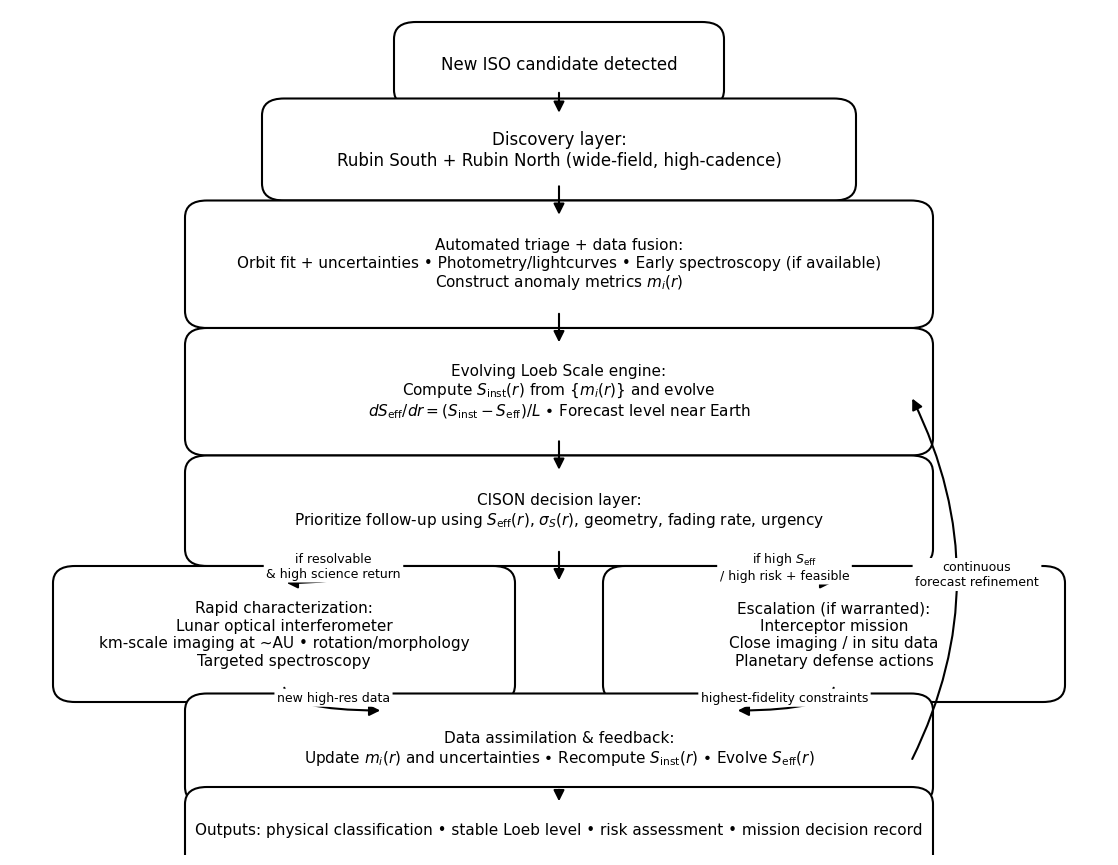}
    \caption{A full summary of how the CISON network would treat a newly identified potential ISO candidate.}
    \label{flow}
\end{figure}

The limitations outlined in the previous section point toward an observational strategy in which no single facility, mission class or observational mode can simultaneously satisfy the requirements of ISO discovery, physical characterization, and risk assessment. Instead a coordinated observational architecture is required in which different components are explicitly optimized for distinct roles and are coupled through rapid information flow and decision logic. 
\\
\\
At the discovery level, the core requirement is maximal sky coverage with high cadence and sufficient depth to detect fast-moving, faint objects over short visibility windows. An observatories setup \cite{vc1thomas2020vera,vc2blum2022snowmass2021,vc3sebag2020vera} which covers both hemispheres is a configuration that naturally satisfies this requirement. Let the instantaneous survey rate be characterized by the $\acute{e}$tendue $\mathcal{E}=A\Omega$, where $A$ is the collecting area and $\Omega$ the field of view. The expected ISO detection rate scales approximately as
\begin{equation}
\Gamma \propto \mathcal{E}\, \epsilon(m_{\rm lim},v,\alpha),
\end{equation}
where $\epsilon$ encodes detection efficiency as a function of limiting magnitude, object velocity and phase geometry. A northern Rubin observatory twin effectively doubles the accessible sky and removes seasonal blind spots, increasing $\Gamma$ not merely by a factor of two but by mitigating cadence gaps that disproportionately affect short-lived ISO visibility windows. If the characteristic visibility time is $t_{\rm vis}\sim \mathcal{O}(1)$ month, then continuous hemispheric coverage increases the probability that an ISO is detected early in its inbound trajectory, thereby increasing the lead time available for follow-up by weeks rather than days. This additional lead time is critical because the information content of subsequent observations scales nonlinearly with distance and brightness.
\\
\\
Discovery alone, however, does not address the dominant inference degeneracies. The second layer of the architecture therefore consists of rapid response, high angular resolution characterization, triggered automatically by discovery alerts and informed by real time orbital and photometric inference. The fundamental quantity controlling the diagnostic power of imaging is the achievable linear resolution $\ell=\Delta\lambda/D$, where $\Delta$ is the object geocentric distance, $\lambda$ the observing wavelength and $D$ the effective baseline. For optical wavelengths $\lambda\sim 0.5\,\mu{\rm m}$ and distances $\Delta\sim 1\,{\rm AU}$, resolving kilometer-scale structure requires effective baselines satisfying equation \eqref{kmscale}, which are inaccessible to terrestrial facilities in an operationally flexible mode due to atmospheric turbulence and scheduling constraints. A lunar based optical interferometer  operating in a vacuum environment with stable thermal and mechanical conditions, naturally reaches this regime as the absence of atmospheric seeing allows diffraction-limited performance, while the lunar surface enables baselines at the required scale. Direct imaging at this resolution collapses multiple degeneracies simultaneously by providing constraints on shape, aspect ratio, binarity and surface structure, thereby breaking the size-albedo-shape degeneracy inherent in unresolved photometry.
\\
\\
The quantitative gain from such imaging can be expressed in terms of parameter inference. In particular, if $\boldsymbol{\theta}$ denotes the set of physical parameters and $\mathbf{d}$ the data, then the Fisher information matrix scales as
\begin{equation}
\mathcal{I}_{ij} \sim \left\langle \frac{\partial \ln p(\mathbf{d}\mid\boldsymbol{\theta})}{\partial \theta_i}\frac{\partial \ln p(\mathbf{d}\mid\boldsymbol{\theta})}{\partial \theta_j}\right\rangle,
\end{equation}
and resolved imaging introduces new observables whose derivatives with respect to $\theta_i$ are orthogonal to those from photometry and astrometry alone. In practice, this means that posterior volumes shrink by orders of magnitude once even coarse spatial resolution is achieved, transforming qualitative speculation into quantitative constraint. Moreover we see that repeated imaging over a short arc allows direct measurement of rotation state and non principal axis motion, which feeds back into improved modeling of non-gravitational accelerations.
\\
\\
The third component of the architecture is the interceptor mission, which occupies a different region of the cost-information space. Interceptors are not discovery instruments but high cost, high information assets capable of in situ measurements and their feasibility depends sensitively on warning time and geometry. If $v_{\rm rel}$ is the relative velocity between the spacecraft and the ISO and $\Delta v$ the achievable propulsion boost, then intercept feasibility requires
\begin{equation}
\Delta v \gtrsim v_{\rm rel}\left(1-\frac{t_{\rm launch}}{t_{\rm encounter}}\right),
\end{equation}
where $t_{\rm launch}$ is the time required to launch the interceptor (which is equal to zero if the interceptor is already in space) and $t_{\rm encounter}$ is the time left before encountering the ISO. This implies that early discovery and rapid orbit determination are prerequisites. The proposed architecture ensures that only a small subset of ISOs, selected on the basis of high scientific return or potential risk are escalated to this level. In this sense, interceptors represent the final rung in a hierarchical response ladder rather than a default solution.
\\
\\
This layered architecture directly resolves the four dominant limitations identified earlier. Cadence and visibility constraints are mitigated by dual hemisphere discovery while  photometric and astrometric degeneracies are broken by spatially resolved imaging. Ambiguities in non-gravitational acceleration are addressed by direct morphology, rotation and mass estimates. The transient nature of ISO information is countered by an explicit rapid response design that minimizes latency between detection and characterization and importantly these cures do not rely on speculative technology but on recombining existing and planned capabilities into a coherent system.
\\
\\
Within this framework, it becomes both logical and strategically compelling to integrate ISO imaging into the Artemis Lunar program \cite{art1creech2022artemis,art2smith2020artemis,art3angelopoulos2011artemis,art4mcintosh2020nasa,art5noble2024implementation,art6sweetser2011artemis,art7figueroa2022risk}. A lunar optical interferometer is already motivated by stellar and exoplanetary science, but ISO imaging imposes requirements that are naturally aligned with the lunar environment. The baseline length required for km-scale resolution at AU distances is of order $10^2$ m (see Eq. \eqref{kmscale}), the pointing requirements are compatible with non sidereal tracking, and the duty cycle is well matched to a target-of-opportunity mode rather than continuous surveying. From an information theoretic perspective we see that the gain from even a single resolved ISO image can exceed that from years of unresolved photometric monitoring, because it constrains parameters that are otherwise unconstrained. If one defines an information gain $\Delta\mathcal{I}$ as the reduction in posterior entropy, then
\begin{equation}
\Delta\mathcal{I} = H[p(\boldsymbol{\theta}\mid\mathbf{d}_{\rm pre})] - H[p(\boldsymbol{\theta}\mid\mathbf{d}_{\rm post})],
\end{equation}
is expected to be dominated by the addition of spatially resolved data and in this sense we see that ISO imaging represents a high-leverage scientific return for a relatively modest incremental cost once a lunar interferometric capability exists.
\\
\\
In summary, a coordinated ISO network consisting of Rubin South and North for discovery, lunar interferometry for rapid high-resolution characterization and interceptors for exceptional cases constitutes a logically consistent, quantitatively justified, and operationally feasible architecture. It directly addresses the structural limitations of current ISO astronomy, rational prioritization based on urgency needs (as explained in the example in the next section with Loeb scale categorization) and provides a compelling scientific justification for incorporating ISO imaging into the broader objectives of lunar exploration. This architecture transforms ISO studies from an opportunistic, discovery driven activity into a mature observational discipline with a clear end-to-end strategy. We would like to name such a network as the "Comprehensive InterStellar Objects Network", or CISON as an abbreviation. The architecture of this network has been summarized in the flow chart of figure \ref{flow}. 
\section{A Hypothetical Case Study: ISO 100I/X and Loeb Scale Evolution within CISON}
To illustrate concretely how CISON transforms ISO science and risk assessment, we consider a hypothetical 100th detected interstellar object which we hereafter denote as 100I/X. We assume that 100I/X is discovered on a hyperbolic inbound trajectory with hyperbolic excess speed $v_\infty \sim 30$ km $s^{-1}$ and a perihelion distance of order $q \sim 0.8 {\rm AU}$, which is broadly representative of the parameter space expected for future ISO detections. The purpose of this section is not to expect the properties of a specific object, but to compare in a controlled and quantitative manner, the information that would be obtained under the current observational paradigm versus that obtained under CISON.
\\
\\
Under the present ISO discovery and follow-up framework, 100I/X would most likely be detected by a wide-field ground based survey several weeks before perihelion, at an apparent magnitude near the survey limit. Its apparent brightness would follow Eq. \eqref{mag} and it would imply a rapidly evolving signal as both heliocentric and geocentric distances change. Initial characterization would consist primarily of sparse astrometry and broadband photometry over a short time baseline $T\sim{\rm weeks}$ and from these data, one would infer a preliminary orbit alongside a poorly constrained absolute magnitude $H$ and a light curve with limited phase coverage. The effective constraints would be dominated by the degeneracy between size, albedo and shape, such that a factor of $\sim 3$-$5$ uncertainty in linear size would be typical. Light-curve amplitudes might suggest elongation, but without resolved imaging the inferred axis ratio would remain model dependent. Spectroscopy would be photon limited and restricted to coarse reflectance slopes and thus providing only weak compositional discrimination. Any detected non-gravitational acceleration $a_{\rm ng}$ would be interpretable either as weak outgassing or radiation pressure, with the ambiguity persisting due to the lack of independent constraints on mass and area-to-mass ratio.
\\
\\
Within the CISON framework, the same object 100I/X would be discovered earlier and with greater sky completeness due to the combined Rubin South and Rubin North coverage. If discovery occurs $\Delta t\sim 30$ days earlier along the inbound trajectory, the distances at first detection could be smaller by a factor of $\sim 1.5$ compared to a single hemisphere scenario and this would be corresponding to a brightness gain from the object's nucleus of
\begin{equation}
\Delta m \sim 5\log_{10}(1.5^2) \sim 1.8,
\end{equation}
which is a factor of $\sim 5$ times higher flux and this alone increases the available photon budget for early characterization by a factor of $\sim 5$. More importantly, early detection enables rapid triggering of CISON follow-up assets as within days of discovery, a lunar optical interferometer could obtain diffraction limited imaging with effective baseline $D\sim 100\,{\rm m}$ at optical wavelengths, providing a linear resolution
\begin{equation}
\ell \sim \frac{\lambda \Delta}{D} \sim 100{\rm m},
\end{equation}
at $\Delta\sim 0.1\,{\rm AU}$. Even a marginally resolved image at this scale would immediately fix the characteristic size, projected shape and presence or absence of binarity. The uncertainty in effective radius would shrink from order unity to the $\sim 10$-$20\%$ level, collapsing the size-albedo degeneracy and anchoring all subsequent physical inference.
\\
\\
Repeated imaging over a span of days would directly reveal the rotation state and any non principal axis motion as well and the moment of inertia ratios inferred from resolved shape would allow dynamical modeling of torques, making it possible to test whether an observed $a_{\rm ng}$ is consistent with plausible outgassing or radiation pressure. In effect, equation \eqref{ng} could be evaluated with independent estimates of $M$ derived from volume and density priors informed by morphology. This would lead to reducing the degeneracy space by orders of magnitude. Spectroscopic follow-up, guided by precise ephemerides and non sidereal tracking, would benefit from higher signal-to-noise due to earlier and closer observations, thus enabling detection limits on volatile species that are inaccessible under the current paradigm.
\\
\\
The cumulative information gain from CISON can be summarized in Bayesian terms better, and for that let $\boldsymbol{\theta}$ represent the physical parameter vector of 100I/X. Under current capabilities, the posterior volume $V_{\rm post}$ remains large because key parameters are weakly constrained but under CISON, the addition of resolved imaging and early spectroscopy introduces new likelihood terms that are nearly orthogonal in parameter space, such that
\begin{equation}
\frac{V_{\rm post}^{\rm CISON}}{V_{\rm post}^{\rm current}} \ll 1,
\end{equation}
with reductions by factors of $\sim 10^2$-$10^3$ plausible for subsets of parameters such as size, shape, and rotation state. This represents not an incremental but a qualitative transition in the kind of science that can be performed.
\\
\\
The advantage of CISON becomes even more pronounced in a hypothetical scenario where 100I/X poses a potential hazard to Earth and for that we need to explicitly incorporate the mathematical structure and differential evolution of the Loeb Scale \cite{eldadi2025loeb,ref1eldadi2025surveys,ref4eldadi2025adaptive,loe1trivedi2025quantitative,loe2trivedi2025evolving}. This allows us to demonstrate not only how CISON improves physical characterization, but also how it enables faster, more stable, and more predictive determination of an object’s effective Loeb level as it approaches Earth. Throughout this we emphasize that the key advance of CISON is not merely higher data quality, but the ability to populate and evolve the anomaly metrics entering the Loeb Scale with significantly reduced uncertainty and latency. We assume that 100I/X is detected at a heliocentric distance $r_{\rm det}\sim 3\,{\rm AU}$ on an inbound hyperbolic trajectory with $v_\infty\sim 30\,{\rm km\,s^{-1}}$. At detection, observational constraints are sparse and each Loeb anomaly metric $m_i(r)$ is weakly constrained, with broad uncertainties reflecting limited astrometry, low signal-to-noise photometry and little or no spectroscopy and so the instantaneous Loeb score at detection \cite{loe1trivedi2025quantitative} is
\begin{equation}
S_{\rm inst}(r_{\rm det}) = \sum_i w_i m_i(r_{\rm det}) + \sum_{i<j} w_{ij} m_i(r_{\rm det}) m_j(r_{\rm det}),
\end{equation}
where $w_i$ are the weights of the anomalies $m_i$. The score is dominated by prior assumptions rather than data and with current ISO capabilities, the dominant contributors at this stage would be the trajectory anomaly $E(r)$ and possibly a shape anomaly $C(r)$ inferred from sparse light curves, while metrics such as the non-gravitational acceleration anomaly $A(r)$, spectral anomaly $B(r)$ and albedo anomaly $D(r)$ would remain poorly constrained. As a result, the uncertainty $\sigma_S$ propagated from
\begin{equation}
\sigma_S^2 \simeq \sum_i \left(w_i + \sum_{j\neq i} w_{ij} m_j\right)^2 \sigma_{m_i}^2,
\end{equation}
would remain large and any inferred Loeb level would be unstable against additional data.
\\
\\
Within the CISON framework, the early discovery enabled by dual-hemisphere Rubin coverage significantly alters this situation and the game changes. Earlier detection increases the number of astrometric epochs and reduces uncertainties in the inbound asymptote, which ends up leading to a more rapid tightening of the trajectory anomaly metric $E(r)$. Because $E(r)$ depends logarithmically on the improbable nature of the arrival geometry, even modest improvements in orbital uncertainty can lead to appreciable changes in
\begin{equation}
E(r) = \mathrm{clamp}\!\left(\frac{-\log_{10} p(r)}{X},0,1\right),
\end{equation}
with $p(r)$ being the isotropic arrival probability. Under current capabilities, $p(r)$ may shrink by orders of magnitude only after weeks of follow-up, whereas under CISON this contraction occurs earlier, allowing $E(r)$ to approach its asymptotic value at substantially larger heliocentric distances. The most dramatic difference, however, arises once lunar interferometric imaging is triggered as part of CISON. Direct imaging of 100I/X at sub-kilometer-scale resolution collapses the uncertainty in the shape anomaly metric
\begin{equation}
C(r) = \mathrm{clamp}\!\left(\frac{\log_{10} R(r)}{\log_{10} R_{\max}},0,1\right),
\end{equation}
by directly measuring the aspect ratio $R$ rather than inferring it from light curve amplitudes subject to viewing-angle degeneracies. In practice, this means that $\sigma_C(r)$ decreases exponentially with the number of resolved images rather than slowly with the number of unresolved light curve points. This rapid reduction in $\sigma_C$ feeds directly into a sharper estimate of $S_{\rm inst}(r)$.
\\
\\
Similarly resolved size measurements combined with photometry allow an independent determination of albedo, dramatically improving the albedo anomaly metric $D(r)$. Instead of relying on population level priors, we see the tail probability entering
\begin{equation}
D(r) = \mathrm{clamp}\!\left(\frac{s_{\rm albedo}(r)}{s_{\rm albedo}(r)+K_D},0,1\right),
\end{equation}
can be computed with a directly measured $p_V$ shrinking $\sigma_D$ by an order of magnitude relative to current practice. In parallel, early and higher signal-to-noise spectroscopy enabled by earlier detection improves the spectral anomaly metric $B(r)$ as well, allowing its radial evolution to be meaningfully modeled rather than remaining noise dominated until near perihelion.
\\
\\
The combined effect of these improvements is most naturally expressed through the differential formulation of the Loeb Scale \cite{loe2trivedi2025evolving}. The effective Loeb score $S_{\rm eff}(r)$ evolves according to the relaxation equation
\begin{equation}
\frac{dS_{\rm eff}}{dr} = \frac{S_{\rm inst}(r) - S_{\rm eff}(r)}{L},
\end{equation}
where $L$ is the characteristic relaxation length and under current observational setups the instantaneous score $S_{\rm inst}(r)$ fluctuates significantly due to measurement noise, forcing one to choose a large $L$ to avoid spurious classification changes. This in turn delays convergence of $S_{\rm eff}(r)$ toward its final value near Earth. However, with CISON, the reduced uncertainties in $m_i(r)$ suppress high frequency fluctuations in $S_{\rm inst}(r)$. This ends up permitting a smaller effective $L$ and therefore faster convergence of $S_{\rm eff}(r)$ toward its asymptotic value.
\\
\\
This distinction has direct operational consequences. By integrating the evolution equation forward to $r_\oplus\simeq 1\,{\rm AU}$, one obtains a predictive distribution for the Earth-encounter score
\begin{equation}
P\!\left(S_{\rm eff}(r_\oplus)\mid D_{\rm det}\right) = \int d\boldsymbol{\theta}\,P(\boldsymbol{\theta}\mid D_{\rm det})\,\delta\!\left(S_{\rm eff}(r_\oplus;\boldsymbol{\theta})-s\right),
\end{equation}
where $\boldsymbol{\theta}$ denotes the metric parameters. Under current capabilities, this distribution remains broad until late times which leads to making early risk assessment unreliable. Under CISON, the early collapse of uncertainties in $C(r)$, $D(r)$, $B(r)$ and $E(r)$ sharply narrows this predictive distribution weeks earlier, which allows a stable forecast of the eventual Loeb level long before Earth encounter.
\\
\\
In a hypothetical hazardous scenario, where 100I/X exhibits a non-gravitational acceleration and a trajectory anomaly suggestive of Earth crossing risk, this distinction becomes critical. The impact-risk metric $H(r)$ enters the composite score linearly and through synergies with other anomalies and because the uncertainty in $H(r)$ scales with the uncertainty in the propagated trajectory, which itself depends on the uncertainty in $A(r)$, faster physical discrimination between outgassing and radiation-pressure scenarios under CISON reduces $\sigma_{A}$ and therefore $\sigma_{H}$. If $\sigma_{A}$ is reduced by an order of magnitude, the resulting uncertainty in the predicted Earth encounter distance scales down proportionally, allowing the Loeb score to cross from critical thresholds on timescales of days rather than months.
\\
\\
Consequently, under current methods the classification of 100I/X might oscillate between Loeb levels as new data arrive, with a definitive hazard assessment possible only close to perihelion or even post-encounter. Under CISON, the evolving effective score $S_{\rm eff}(r)$ would stabilize early enabling rapid declaration of whether the object is benign or warrants escalation to an interceptor mission. This case study highlights a central conceptual point, that being that CISON does not merely improve ISO observations in an incremental sense, but fundamentally alters the behavior of the Loeb Scale as a dynamical classifier. By enabling earlier, more accurate determination of the individual anomaly metrics and by suppressing spurious fluctuations in the instantaneous score, CISON accelerates the convergence of the effective Loeb score and transforms it into a genuinely predictive quantity. In doing so, it allows ISO science and planetary defense to transition from reactive classification near Earth to anticipatory decision making based on stable forecasts, which is fully reliant on the mathematical intent of the evolving Loeb Scale framework.

\section{Conclusions}

In this work we have argued that current studies of interstellar objects are limited not by the absence of discovery facilities, but by the lack of a coherent end-to-end observational architecture that links discovery, characterization and decision making. By identifying the dominant structural issues in present ISO studies and formulating a coordinated response, we have proposed the Comprehensive ISO Network (CISON) as a physically motivated and operationally feasible framework. CISON separates discovery and characterization into complementary layers, combining dual-hemisphere Rubin observatory with rapid  response high resolution follow-up and selective escalation to interceptor missions. It directly addresses the core limitations of cadence, degeneracy and perishable information that limit existing studies.
\\
\\
A central result of this work is that such an architecture fundamentally alters not only the quantity but the quality of information available for newly discovered ISOs. Through early detection, spatially resolved imaging and rapid physical discrimination of non-gravitational effects, CISON enables decisive collapse of parameter degeneracies that otherwise persist until late times. When coupled to the differential formulation of the Loeb Scale \cite{eldadi2025loeb,ref1eldadi2025surveys,ref4eldadi2025adaptive,loe1trivedi2025quantitative,loe2trivedi2025evolving}, this improvement translates into faster, more stable and genuinely predictive classification of interstellar objects. The evolving Loeb score \cite{loe2trivedi2025evolving} becomes an operational diagnostic rather than a retrospective label, allowing risk assessment and scientific prioritization to proceed on timescales of days to weeks instead of months.
\\
\\
More broadly, CISON reframes ISO astronomy as a mature, anticipatory discipline rather than an opportunistic byproduct of time domain surveys as it is right now. By naturally motivating the inclusion of ISO imaging within lunar infrastructure such as the Artemis lunar program, the proposed architecture embeds interstellar science within the long term expansion of observational capabilities beyond Earth. In doing so, it establishes a template for how future astronomical frontiers may be explored, that being through tightly integrated networks that combine wide-field discovery, precision characterization and quantitative decision frameworks. As ISO detection rates rise in the coming decades, such an approach will be essential not only for maximizing scientific return but also for responsibly assessing rare objects that may carry profound implications for planetary defense, technosignature searches and our understanding of our cosmic environment.
\section*{Acknowledgements}
The work of OT was supported in part by the Vanderbilt Discovery Alliance Fellowship. The work of AL was supported in part by Harvard's Galileo Project and the Black Hole Initiative, which is funded by GBMF and JTF.

\bibliography{references}
\bibliographystyle{unsrt}
\end{document}